\documentclass[twocolumn,showpacs,superscriptaddress,preprintnumbers,amsmath,amssymb]{revtex4}
\usepackage{graphicx}
\usepackage{dcolumn}
\usepackage{bm}
\usepackage{subfigure}

\begin{document}

\preprint{}

\title{Equilibration and relaxation times at the chiral phase transition including reheating}

\author{Marlene Nahrgang}
\affiliation{Institut f\"ur Theoretische Physik, Goethe-Universit\"at, Max-von-Laue-Str.~1, 
60438 Frankfurt am Main, Germany}
\affiliation{Frankfurt Institute for Advanced Studies (FIAS), Ruth-Moufang-Str.~1, 60438 Frankfurt am Main, Germany}

 \author{Stefan Leupold}
 \affiliation{Department of Physics and Astronomy, Uppsala University, Box 516, 75120 Uppsala, Sweden}

 \author{Marcus Bleicher}
 \affiliation{Institut f\"ur Theoretische Physik, Goethe-Universit\"at, Max-von-Laue-Str.~1,
 60438 Frankfurt am Main, Germany}
\affiliation{Frankfurt Institute for Advanced Studies (FIAS), Ruth-Moufang-Str.~1, 60438 Frankfurt am Main, Germany}

\date{\today}

\begin{abstract}
We investigate the relaxational dynamics of the order parameter of chiral symmetry breaking, the sigma mean-field, with a heat bath consisting of quarks and antiquarks. A semiclassical stochastic Langevin equation of motion is obtained from the linear sigma model with constituent quarks. The equilibration of the system is studied for a first order phase transition and a critical point, where a different behavior is found. At the first order phase transition we observe the phase coexistence and at a critical point the phenomenon of critical slowing down with large relaxation times. We go beyond existing Langevin studies and include reheating of the heat bath by determining the energy dissipation during the relaxational process. The energy of the entire system is conserved. In a critical point scenario we again observe critical slowing down. 
\end{abstract}

\maketitle
\section{Introduction}\label{chap:equilibration}

In heavy-ion collisions matter is created under extreme conditions, in small systems and with fast dynamics. It is, therefore, likely that nonequilibrium effects play an important role in the evolution of the fireball. Such effects are of particular importance at the conjectured critical point in the QCD phase diagram. Signals of the critical point are based on the growth of the correlation, which leads to large fluctuations \cite{Stephanov:1998dy,Stephanov:1999zu}. Naturally the size of the system itself is a limit to the correlation length. Here, the renormalization group methods of finite size scaling could be applied to localize the critical point \cite{Palhares:2010zz}.
Finite time effects due to the phenomena of critical slowing down turn out to be more limiting on the growth of the correlation length than finite size effects \cite{Berdnikov:1999ph}. The relaxation times at a critical point become infinitely long. Even if the system is in thermal equilibrium above $T_c$ it is necessarily driven out of equilibrium when crossing the critical point.

For the dynamics in the vicinity of a phase transition or rapid crossover the study of an order parameter is of particular interest. For a proper description of nonequilibrium effects the following ingredients are necessary:
\begin{itemize}
 \item  a (semiclassical) equation of motion for the long-wavelength modes related to the order parameter, which includes damping and noise for the long-wavelength modes, i.e.\ a Langevin dynamics needs to be formulated (see e.g. \cite{Greiner:1998vd} and references therein).;
 \item  a fluid dynamic description of the rest of the system (``heat bath'') \cite{Mishustin:1998yc,Paech:2003fe};
\item the back reaction of the long-wavelength modes on the heat bath should be consistently included.
\end{itemize}
This latter aspect is important because for a fireball with an extension of at most a few femtometers the ``heat bath'' cannot be regarded as infinitely large. But it is missing in most of the Langevin descriptions \cite{Greiner:1996dx,Biro:1997va,Rischke:1998qy,CassolSeewald:2007ru} while chiral fluid dynamic models typically disregard the Langevin dynamics \cite{Mishustin:1998yc,Paech:2003fe}.

In \cite{Nahrgang:2011mg} we have derived a Langevin-type equation of motion for a sigma field coupled to a heat bath of quarks. We have determined explicit expressions for the damping coefficient and the correlation of the noise term. Within the formalism of the two-particle irreducible (2PI) effective action  \cite{Luttinger:1960ua,Lee:1960zza,Baym:1961zz,Baym:1962sx,Cornwall:1974vz,Ivanov:1998nv,vanHees:2001ik,vanHees:2001pf,vanHees:2002bv} we were able to obtain the equilibrium properties of the heat bath consistently. During the relaxational process of the sigma field energy dissipates from the field to the heat bath, which is thereby reheated. Usually this reheating effect is not taken into account. In a small and dynamic system this back reaction on the heat bath can, however, be important and change the overall evolution of the system.

In this paper we want to investigate the relaxational dynamics of the sigma field and time scales associated with it.
The relaxational properties of the sigma field are important for studies of the phase transition. In a nonequilibrium situation one expects to see the coexistence phase for a first order phase transition, and due to critical slowing down one also expects long relaxation times at the critical point. 

In section \ref{sec:equi_numimplLangevin} we explain the dynamics for the sigma mean-field and the local equilibrium properties of the quarks on the basis of the linear sigma model with constituent quarks. We study the equilibration properties of the sigma field within our model for a static and isothermal heat bath in section \ref{sec:equi_isothermal}. During the relaxation process the sigma field loses energy due to dissipation. This is studied in comparison with the energy conservation of the entire system in section \ref{sec:equi_Edissipation}. Finally, we include this energy exchange and take reheating of the heat bath into account in section \ref{sec:equi_reheating}.

\section{The dynamics of the sigma field}\label{sec:equi_numimplLangevin}
In our approach \cite{Nahrgang:2011mg} the relaxational dynamics of the order parameter of chiral symmetry breaking, the sigma mean-field, is given by the linear sigma model with constituent quarks \cite{GellMann:1960np}. The quark degrees of freedom are assumed to be in local thermal equilibrium, such that they constitute the heat bath. The sigma mean-field is propagated by a stochastic Langevin equation of motion.

\subsection{The linear sigma model with constituent quarks}
The linear sigma model with constituent quarks describes the chiral phase transition. Here, the $\sigma$ and $\pi$ mesons couple to quarks. The Lagrangian reads 
\begin{equation}
\begin{split}
{\cal L}&=\bar{q}[i\gamma^\mu\partial_\mu-g(\sigma+i\gamma_5\vec\tau{\vec\pi})]q \\
  &\quad+ \frac{1}{2}(\partial_\mu\sigma\partial^\mu\sigma)+
 \frac{1}{2}(\partial_\mu\vec{\pi}\partial^\mu\vec{\pi}) 
- U(\sigma, \vec{\pi}) \, ,
\label{eq:LGML}
 \end{split}
\end{equation}
with the constituent quark field $q=\left(u,d\right)$, the sigma field $\sigma$ and the pion fields $\vec{\pi}$. $g$ is the coupling strength of the interaction between the chiral fields and the quarks.  $\vec{\tau}$ are the isospin Pauli matrices. The classical chiral potential is given by
\begin{equation}
U\left(\sigma, \vec{\pi}\right)=\frac{\lambda^2}{4}\left(\sigma^2+\vec{\pi}^2-\nu^2\right)^2-h_q\sigma-U_0\, .
\label{eq:Uchi}
\end{equation} 
The Lagrangian (\ref{eq:LGML}) is invariant under transformations of the chiral group ${\rm SU}_{\rm L}(2)\times{\rm SU}_{\rm R}(2)$ if the explicit symmetry breaking term $h_q$ vanishes in the potential (\ref{eq:Uchi}).
The parameters in (\ref{eq:Uchi}) are chosen such that the vacuum expectation values are reproduced, where $\langle\sigma\rangle=f_\pi=93$~MeV and $\langle\vec\pi\rangle=0$. Here, chiral symmetry is spontaneously broken. The explicit symmetry breaking term is $h_q=f_\pi m_\pi^2$ with the pion mass $m_\pi=138$~MeV. Then, $\nu^2=f_\pi^2-m_\pi^2/\lambda^2$. A realistic vacuum sigma mass $m_\sigma^2=2\lambda^2 f_\pi^2 + m_\pi^2\approx 604$~MeV is given for  $\lambda^2=20$. In order to have zero potential energy in the ground state the term $U_0=m_\pi^4/(4\lambda^2)-f_\pi^2 m_\pi^2$ is subtracted. At a coupling $g=3.3$ the constituent quark mass in vacuum is $m_q=306.9$~MeV.

The Lagrangian (\ref{eq:LGML}) treats the quarks and antiquarks and the mesons on equal footing. In the real world confining forces recombine quarks and antiquarks in mesons and baryons below the confinement critical temperature. The aspect of confinement is not included in the linear sigma model with constituent quarks. We can, thus, investigate the pure effect of the chiral phase transition. For extensions of the model gluons can be included on the level of the dilaton field \cite{Mishustin:1998yc} or the Polyakov loop \cite{Schaefer:2007pw}.

\begin{figure}[t]
 \centering
 \includegraphics[width=0.45\textwidth]{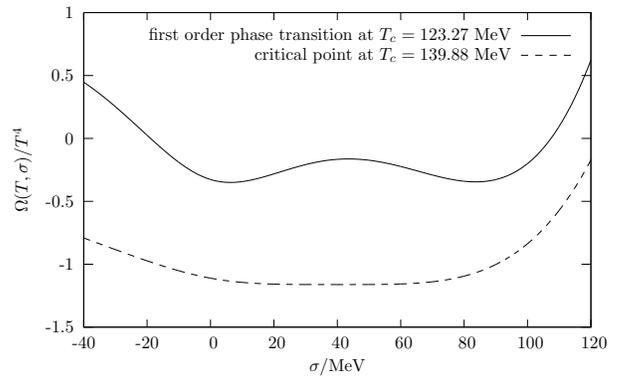}
 \caption{The effective potential shown for the first order phase transition at a coupling $g=5.5$ and the transition temperature $T_c=23.27$~MeV and the critical point at a coupling $g=3.3$ and the transition temperature $T_c=139.88$~MeV. At the first order phase transition the effective potential has two minima, one close to vanishing $\sigma$ and one close to the vacuum expectation value $\langle\sigma\rangle=f_\pi$. At the critical point the minimum is very flat around the minimum. The effective potential is given by the thermodynamic potential to be calculated in (\ref{eq:thermodynamicpot}).}
 \label{fig:veffomega}
\end{figure}

The linear sigma model with constituent quarks exhibits the full spectrum of the suggested chiral phase structure of QCD \cite{Friman:2011zz}. It has a crossover transition at vanishing $\mu_B$ and a first order transition line at high $\mu_B$ and lower temperatures, which terminates in a critical point at $T_c$ \cite{Scavenius:2000qd}. The phase transition can also be tuned by fixing $\mu_B=0$ and changing the coupling constant $g$ \cite{Scavenius:2000bb,Aguiar:2003pp}. At large couplings the phase transition is of first order. Here, the effective potential has two minima, which are separated by a barrier. Lowering the coupling constant one finds a critical point, where the curvature at the minimum of the effective potential is very flat. In this work we use $g=5.5$ for a first order phase transition with the transition temperature $T_c=123.27$~MeV and $g=3.63$ for a critical point with the transition temperature $T_c=139.88$~MeV. For both cases the effective potential is shown in figure \ref{fig:veffomega}. The purpose of the present work is a first qualitative assessment of the relaxational dynamics of the order parameter consistently coupled to a heat bath of finite size. Therefore we did neither include finite baryo-chemical potential nor pion degrees of freedom. Consequently we have to live with the somewhat unrealistic values of the transition temperature.

\subsection{Local equilibrium properties of the quarks}
Within the formalism of the 2PI effective action \cite{Luttinger:1960ua, Lee:1960zza, Baym:1961zz, Baym:1962sx,Cornwall:1974vz,Ivanov:1998nv, vanHees:2001ik,vanHees:2001pf,vanHees:2002bv}  the local thermodynamics of the quarks is given consistently by the one-loop thermodynamic potential in mean-field approximation.
At $\mu_B=0$ the thermodynamic potential is
\begin{equation}
  \begin{split}
  \Omega(T,\sigma, \vec{\pi})&=U\left(\sigma, \vec{\pi}\right)+\Omega_{q\bar q}(T,\sigma, \vec{\pi})\\
		&=U\left(\sigma, \vec{\pi}\right)-2d_q T\! \int\!\!\frac{{\rm d}^3p}{(2\pi)^3}\ln\left(\!1+\exp\left(\!-\frac{E}{T}\right)\!\right)\, ,
  \end{split}
\label{eq:thermodynamicpot}
\end{equation}
where $d_q=12$ is the degeneracy factor of the quarks for $N_f=2$ flavors, $N_c=3$ colors and the two spin states. The quark mass is generated by nonvanishing expectation values of the chiral fields due to spontaneous symmetry breaking. In the course of the evaluation of the functional determinant in Dirac and isospin space one generates a term defined as the effective mass of the quarks
\begin{equation}
 m_q^2=g^2(\sigma^2+\vec\pi^2)\, .
\end{equation}
Then, the energy of the quarks and antiquarks is
\begin{equation}
 E=\sqrt{\vec p^2+m_q^2}=\sqrt{\vec p^2+g^2(\sigma^2+\vec\pi^2)}\, .
\end{equation}
According to the local value of the temperature and the sigma mean-field the local pressure of the quarks is  
\begin{equation}
   p(\sigma, \vec{\pi},T)= -\Omega_{q\bar q}(T,\sigma, \vec{\pi})\, .\\
\label{eq:lsm_eos1}
\end{equation}
And the local energy density of the heat bath of quarks is
\begin{equation}
\begin{split}
  e(\sigma, \vec{\pi},T)=& T\frac{\partial p(\sigma, \vec{\pi},T)}{\partial T}-p(\sigma, \vec{\pi},T)\\
	   =& 2d_q\int\frac{{\rm d}^3p}{(2\pi)^3}En_{\rm F}(p)\, .
\end{split}
\label{eq:lsm_eos2}
\end{equation}

\subsection{Langevin equation of the sigma field}
The Langevin equation of motion of the sigma field can consistently be derived in the 2PI effective action formalism or, alternatively, in the influence functional method \cite{Feynman:1963fq,Greiner:1998vd}. It gives a reduced description of the entire system with focus on the evolution of the relevant variables, which are propagated explicitly. The details of the environment are eliminated by integrating out the environmental fields in a path integral over the closed time path contour \cite{Schwinger:1960qe,Keldysh:1964ud}. This method was widely used within the $\phi^4$ theory  \cite{Morikawa:1986rp,Gleiser:1993ea,Boyanovsky:1996xx,Greiner:1996dx}, in gauge theories \cite{Bodeker:1995pp,Son:1997qj} and in ${\cal O}(N)$ chiral models \cite{Rischke:1998qy}.

For the linear sigma model with constituent quarks we assume the following splitting: the irrelevant degrees of freedom are the quarks and antiquarks, which constitute the  heat bath, and the relevant sector is that of the sigma field, which we propagate explicitly. 

The Langevin equation for the sigma mean-field reads 
\begin{equation}
 \partial_\mu\partial^\mu\sigma(t,{\bf x})+\frac{\delta U}{\delta\sigma}+\frac{\delta \Omega_{q\bar q}}{\delta\sigma}+\eta(T)\partial_t \sigma(t,{\bf x})=\xi(t,{\bf x})\, .
\label{eq:equi_langevineq}
\end{equation}
We calculated the damping coefficient and the noise correlator in \cite{Nahrgang:2011mg}

For the kinematic range, $m_\sigma(T)>2m_q(T)=2g\sigma_{\rm eq}(T)$, where the decay of the sigma in a quark-antiquark pair is allowed the damping coefficient $\eta$ for the zero mode of the sigma mean-field reads
\begin{equation}
 \eta=g^2\frac{d_q}{\pi}\left(1-2n_{\rm F}\left(\frac{m_\sigma}{2}\right)\right)\frac{1}{m_\sigma^2}\left(\frac{m_\sigma^2}{4}-m_q^2\right)^{3/2}\, .
\label{eq:equi_damping1}
\end{equation}
For the present calculations we use the respective equilibrium values of the sigma mass $m_{\sigma_{\rm eq}}^2(T)=(\partial^2\Omega(T, \sigma)/\partial\sigma^2)|_{\sigma=\sigma_{\rm eq}}$ and $\sigma_{\rm eq}(T)$ to evaluate this criterion. 
We use (\ref{eq:equi_damping1}) also as a approximation for the non-zero modes in the Langevin equation (\ref{eq:equi_langevineq}).

In this framework there would be no damping in the low-temperature phase because the quarks are not light enough to allow for the decay $\sigma \to \bar q q$. Physically this makes sense, because the quarks should be confined anyway. Additional damping is provided by the decay $\sigma \to 2 \pi$. Strictly speaking this is not included in our present framework. Nonetheless, to obtain a more realistic setup we include the zero-temperature damping coefficient from \cite{Rischke:1998qy} when the constituent quark mass is too large for the $\sigma\to\bar qq$ decay. Then for $m_\sigma(T)<2m_q(T)=2g\sigma_{\rm eq}(T)$
\begin{equation}
    \eta=3/{\rm fm}\, .
\label{eq:equi_damping2}
\end{equation}

The stochastic field in the Langevin equation (\ref{eq:equi_langevineq}) has a vanishing expectation value
\begin{equation}
 \langle\xi(t)\rangle_\xi=0\, ,
\end{equation}
and the noise correlation is given by the dissipation-fluctuation theorem
\begin{equation}
 \langle\xi(t)\xi(t')\rangle_\xi=\frac{1}{V}\delta(t-t')m_\sigma\eta\coth\left(\frac{m_\sigma}{2T}\right)\, .
\label{eq:sc_noisecorrelation}
\end{equation}

In this paper we investigate the time evolution of the following quantities.
The volume average of the sigma field for one configuration of the noise field $\xi_{ijk}$ is
\begin{equation}
 \langle\sigma\rangle_n=\frac{1}{N^3}\sum_{ijk}\sigma_{ijk,n}\, ,
\end{equation}
where $N$ is the number of cells in each direction. We average over several different configurations of the noise
\begin{equation}
 \overline{\langle\sigma\rangle}=\frac{1}{N_r}\sum_{n=1}^{N_r}\langle\sigma\rangle_n\, ,
\end{equation}
typically between $N_r=5$ and $N_r=20$ depending on how different the trajectories really are for the various temperatures.

For the numerical implementation of the Langevin equation we apply the well tested algorithm used in \cite{CassolSeewald:2007ru}.
In the calculations presented here, we choose $N=32$. The size of time steps is $\Delta t=0.02$~fm and the lattice spacing is $\Delta x=0.2$~fm.

\section{Equilibration for a global, isothermal heat bath}\label{sec:equi_isothermal}
First, we study the equilibration of the sigma field with a global, i.e space-homogeneous, and isothermal heat bath. Concerning the energy exchange the back reaction of the sigma field on the heat bath is ignored. Therefore, the temperature of the heat bath is constant and determines the shape of the effective potential. This is very different for the first order phase transition and the critical point. We, therefore, expect a different evolution of the sigma field for these two scenarios.

While the initial conditions for the sigma field are varied in the next sections, the time derivative of the sigma field $\partial_t\sigma$ is initially zero. There is no clear physical motivation for the choice of the initial $\partial_t\sigma$. When it is initialized in direction of the relaxation process the sigma field relaxes faster. A random distribution for the initial  $\partial_t\sigma$ averages out.

\subsection{First order phase transition}
The evolution in a  first order scenario is especially interesting because the effective potential has two minima in the spinodal region $108\,{\rm MeV}\simeq T_{\rm sp}^{(1)}<T_c<T_{\rm sp}^{(2)}\simeq128$~MeV, see figure \ref{fig:veffomega} for the effective potential at the transition temperature $T_c=123.27$~MeV. We expect that for some configurations the sigma field relaxes partly to the unstable minimum instead of the true thermal expectation value. This is even more likely in the vicinity of $T_c$, where both minima are almost degenerate. 

In order to study the relaxation of the sigma field to its thermal equilibrium state at the temperature of the heat bath we need a clear nonequilibrium initial situation.
In our investigation we distinguish between two cases for which the initial nonequilibrium situation is realized differently: the equilibration at temperatures above and at temperatures below the transition temperature.

\subsubsection{Equilibration at high temperatures $T>T_c$}
For the equilibration at temperatures above the phase transition temperature we initially distribute the sigma field linearly between $\sigma_{\rm min}\simeq0$ and $\sigma_{\rm max}>0$ such that the initial average of the sigma field is $\langle\sigma\rangle_n\simeq 50$~MeV. The flat distribution is far from being thermal. 

\begin{figure}
 \centering
 \includegraphics[width=0.45\textwidth]{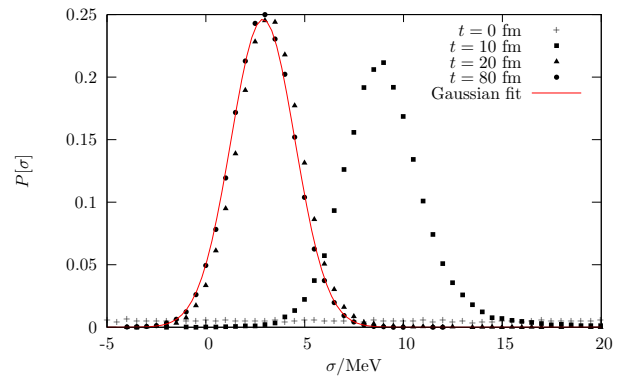}
 \caption{The distribution of the sigma field for $T=160$~MeV at different times. The initial distribution is flat. It quickly turns into a Gaussian centered at a mean of $2.89$~MeV with a width of $1.62$~MeV. The Gaussian fit is to the distribution at $t=80$~fm} 
\label{fig:equi_sigmadis160}
\end{figure}

\begin{figure}
 \centering
 \includegraphics[width=0.45\textwidth]{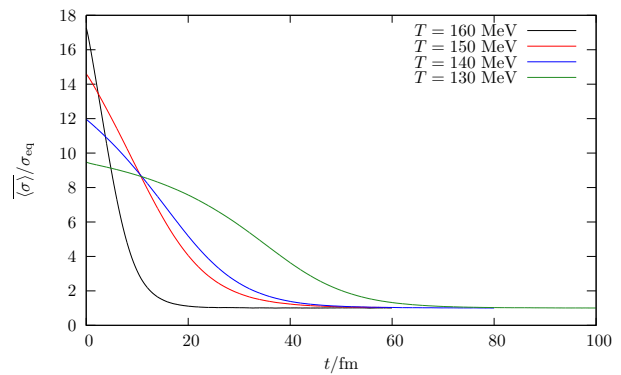}
\caption[Average of the sigma field in a first order phase transition for a static heat bath and temperatures $T>T_{\rm sp}^{(2)}$]{The time evolution of the scaled noise average of the sigma field in a first order scenario for temperatures above the upper spinodal temperature $T>T_{\rm sp}^{(2)}$.}
 \label{fig:equi_avsabovefo}
\end{figure}

In figure \ref{fig:equi_sigmadis160} we see how the initially flat distribution develops for an example evolution at $T=160$~MeV. It becomes Gaussian after times $t\simeq 20$~fm, which correspond to the relaxation times in figure \ref{fig:equi_avsabovefo}. It shows the equilibration of the sigma field to its proper equilibrium value at the respective temperature. This is ensured by the fluctuation-dissipation theorem (\ref{eq:sc_noisecorrelation}), which relates the variance of the noise field to the damping coefficient.

\begin{figure}
 \centering
 \includegraphics[width=0.45\textwidth]{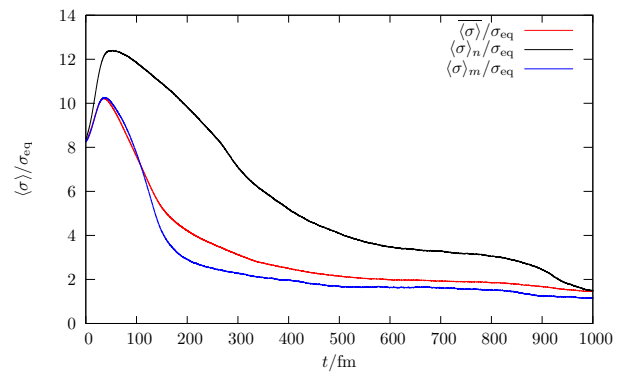}
 \caption{The long relaxation process of the sigma field for $T=125$~MeV. The scaled noise average $\overline{\langle\sigma\rangle}/\sigma_{\rm eq}$ and two individual noise configurations are given. Due to the barrier between the two minima the relaxation to the global minimum is very slow. The individual noise configurations differ substantially.} 
\label{fig:equi_sigmaav125}
\end{figure}

The dynamics of the system becomes different for $T_{\rm sp}>T>T_c$. In figure \ref{fig:equi_sigmaav125} the scaled noise average $\overline{\langle\sigma\rangle}/\sigma_{\rm eq}$ and the volume average $\langle\sigma\rangle_n/\sigma_{\rm eq}$ for two individual noise realizations are shown for $T=125$~MeV. Because of the two minima the system takes a long time to relax to the global minimum at $\sigma_{\rm eq}\simeq6$~MeV. Since the initial average is slightly shifted towards the low-temperature minimum the system first tends to this phase, but finally relaxes to the true minimum at very large times.

\begin{figure}
 \centering
 \includegraphics[width=0.45\textwidth]{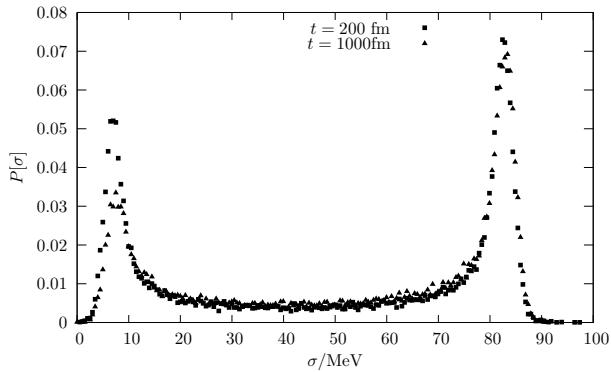}
 \caption{The distribution of the sigma field at the transition temperature $T_c=123.27$~MeV of the first order phase transition scenario. The system shows phase coexistence.}
 \label{fig:equi_sigmadis123}
\end{figure}

At the transition temperature $T_c=123.27$~MeV the distribution of the sigma field for one noise configuration is shown in figure  \ref{fig:equi_sigmadis123}. The system is in the expected phase coexistence and the sigma field does not relax at all. Instead we observe that the system is split into one part in the high-temperature and one part in the low-temperature minimum.

\subsubsection{Equilibration for a quench to temperatures $T<T_c$}
After initializing the sigma field in equilibrium with an initial temperature of $T_{\rm ini}=160\text{ MeV}>T_c=123.27$~MeV the system is suddenly quenched to different temperatures $T<T_{\rm c}$  for which the Langevin equation for the sigma field is solved.

\begin{figure}
 \centering
 \includegraphics[width=0.45\textwidth]{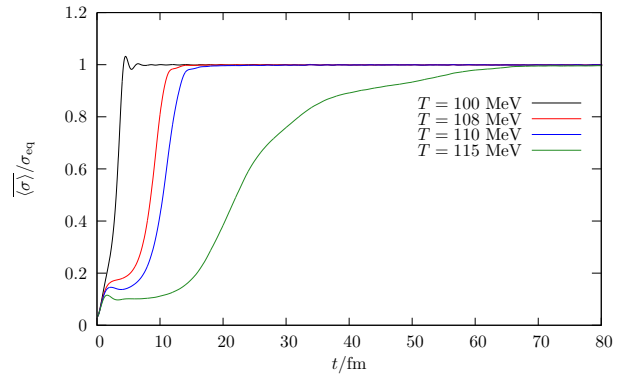}
\caption{The time evolution of the scaled noise average of the sigma field in different quenched scenarios for a first order phase transition scenario.}
 \label{fig:equi_avsbelowfo}
\end{figure}

It is known that for the linear sigma model with constituent quarks the nucleation rates are rather low and that the main relaxation mechanism is that of spinodal decomposition \cite{Scavenius:2000bb}. This is also observed in our calculations. In figure \ref{fig:equi_avsbelowfo} we show the time evolution of the relaxation of the sigma field. For $T=115$~MeV, where we still have a substantial barrier between the two minima, the relaxation times are significantly larger than for lower temperatures. Here, the system remains in the local minimum $\sigma\simeq10$~MeV until at $t\simeq 15$~fm it begins to decay to the global minimum, which is also a slow process. It resembles the case of an exponentially damped system, which decays without oscillations. At $T=100$~MeV we can clearly see the oscillating relaxation. It occurs much faster.

\subsection{Critical point}

\begin{figure}
 \centering
 \includegraphics[width=0.45\textwidth]{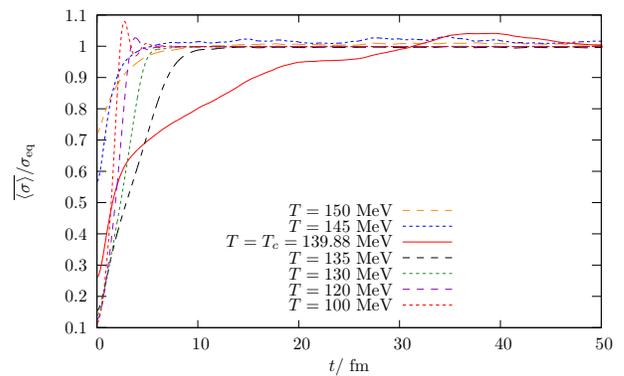}
\caption{The time evolution of the scaled noise average of the sigma field for a critical point scenario for various temperatures. The relaxation time becomes large at the critical point.}
 \label{fig:equi_avscp}
\end{figure}

The effective potential for a scenario with a critical point has only one minimum at all temperatures. It continuously shifts from $\sigma\simeq0$ to the vacuum expectation value $\langle\sigma\rangle_{\rm vac}=f_\pi$ when lowering the temperature. At the critical point this minimum becomes very flat, see figure \ref{fig:veffomega}, and we expect long relaxation times. For different temperatures, the evolution of the noise averaged sigma field can be best compared by choosing the same initial conditions, the thermal equilibrium state at $T=160$~MeV. The system is then quenched to temperatures $T<160$~MeV. The results are shown in figure \ref{fig:equi_avscp}. For low temperatures the potential is steeper and the relaxation process occurs faster. The field oscillates before relaxing. Approaching the critical temperature relaxation times become larger with a clear maximum at $T_c$.

\section{The energy dissipation}\label{sec:equi_Edissipation}
During the relaxation of the sigma mean-field to its equilibrium value the dissipative term in the Langevin equation (\ref{eq:equi_langevineq}) causes energy dissipation. By the interaction with the quarks it is transfered to the heat bath. In \cite{Nahrgang:2011mg} we derived a conserved energy-momentum tensor of the entire system including a dynamics of the heat bath. In this paper, we do not include the fluid dynamic expansion of the heat bath. In this section we investigate the relevant energy exchange between the sigma field and the heat bath. The energy dissipation of the field to the heat bath can be obtained from the energy-momentum tensor of the sigma field
\begin{equation}
 \partial_\mu T_\sigma^{\mu0}=-(g\rho_s+\eta\partial_t\sigma) \partial_t\sigma ,
\label{eq:chfl1_sourceterm}
\end{equation}
where $T_\sigma^{\mu\nu}$ is the energy-momentum tensor of the purely mesonic Lagrangian
 \begin{equation}
 {\cal L}_\sigma=\frac{1}{2}\partial_\mu\sigma\partial^\mu\sigma-U(\sigma)\; .
 \end{equation}
Then, the energy dissipation is described by
\begin{equation}
  \Delta E_{\rm diss}=(g\rho_s+\eta\partial_t\sigma)\partial_t\sigma\Delta t\, .
\label{eq:equi_s0}
\end{equation} 
The total energy of the sigma field is given by
\begin{equation}
 E_{\sigma}=\frac{1}{2}\partial_t\sigma^2+\frac{1}{2}\vec{\nabla}\sigma^2+U(\sigma)\, .
\label{eq:cf_etotsigma}
\end{equation}

It has a kinetic, potential and fluctuation energy term. During relaxation to the vacuum expectation value the potential energy is transfered to kinetic energy as $\partial_t\sigma$ grows. Then, the damping becomes substantial and causes energy dissipation. This flow of energy from the field to the heat bath is given by (\ref{eq:equi_s0}). There is a reverse flow of energy from the heat bath to the field $\Delta E_\xi$ associated with the noise field $\xi$ in the Langevin equation (\ref{eq:equi_langevineq}), which is an averaged quantity balancing the energy dissipation $\Delta E_{\rm diss}$ in equilibrium and thus restoring the proper thermal equilibrium. This was already discussed in \cite{Nahrgang:2011mg}. Assuming that the made approximations in \cite{Nahrgang:2011mg} cause only a small violation of energy conservation we can determine $\Delta E_\xi$ from comparing $\Delta E_{\rm diss}$ to the energy difference in the field before and after each numerical time step $\Delta E_\sigma$.

\begin{figure}
 \centering
 \subfigure[]{\label{fig:equi_sourcetermfo}\includegraphics[width=0.45\textwidth]{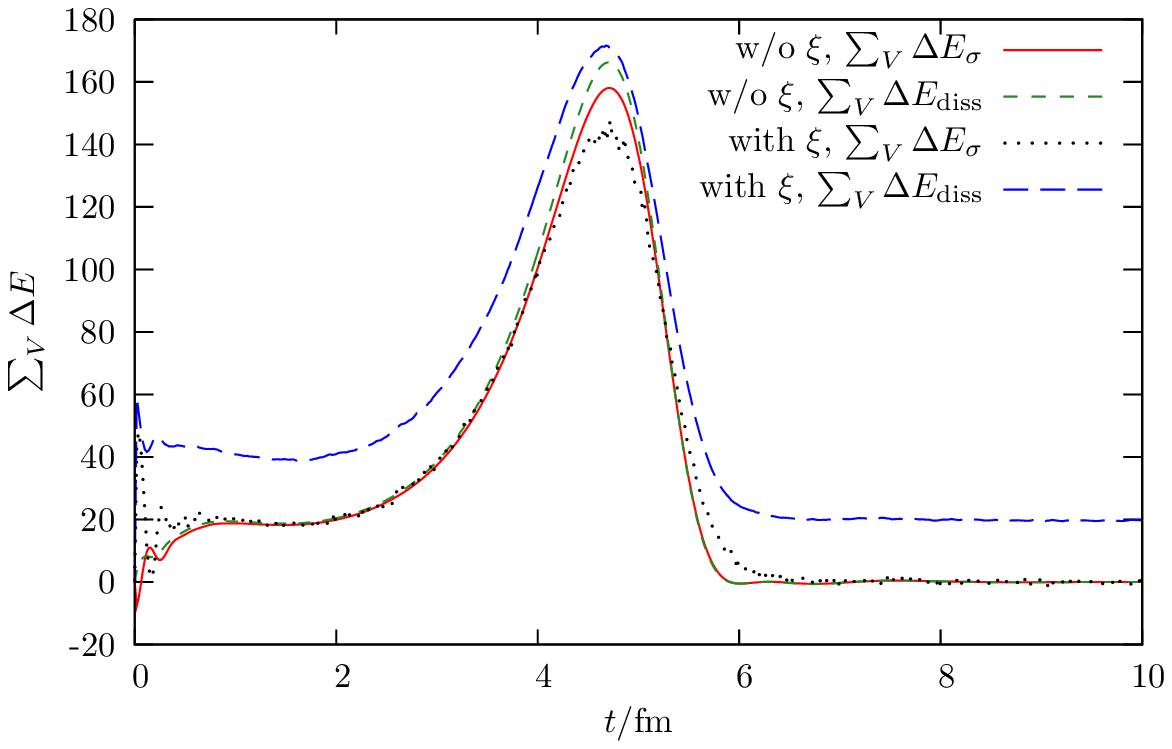}}
 \subfigure[]{\label{fig:equi_sourcetermcp}\includegraphics[width=0.45\textwidth]{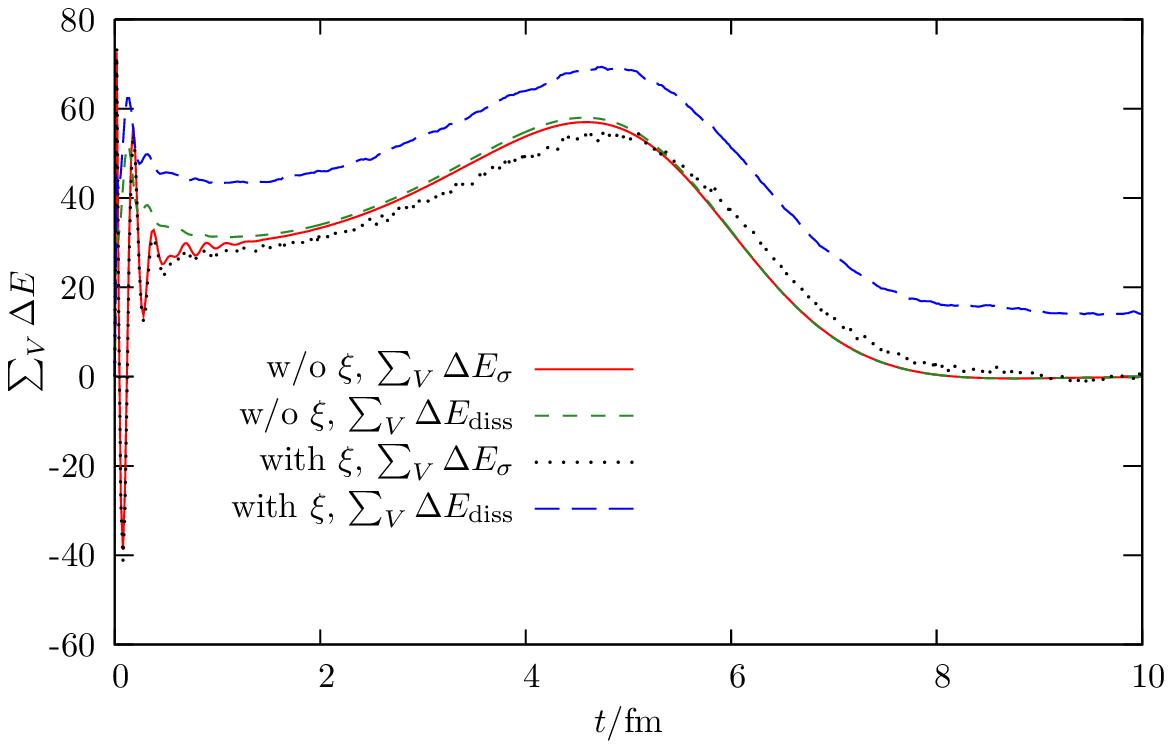}}
\caption{Energy dissipation for a scenario with a first order phase transition \subref{fig:equi_sourcetermfo} and for a critical point scenario \subref{fig:equi_sourcetermcp}. The system is quenched from $T_{\rm ini}=160$ MeV to $T=100$ MeV in the first order phase transition scenario and from $T_{\rm ini}=160$~MeV to $T=130$~MeV in the critical point scenario. The Langevin equation is once solved with the noise term $\xi$ and once without it. For each case the comparison between the total energy dissipation $\sum_V\Delta E_{\rm diss}$ and the energy difference in the field $\sum_V\Delta E_\sigma$ summed over the whole volume is shown.}
 \label{fig:equi_sourcetermtot}
\end{figure}

Here, we first show that the difference between $\Delta  E_{\rm diss}$ and $\Delta E_\sigma$ is small if one ignores the noise term in the Langevin equation. This is shown in figure \ref{fig:equi_sourcetermfo} for the quench from $T_{\rm ini}=160$~MeV to $T=100$~MeV in a scenario with a first order phase transition and for a critical point scenario quenched from $T_{\rm ini}=160$~MeV to $T=130$~MeV in figure \ref{fig:equi_sourcetermcp}. The resulting difference is a measure of the violation of energy conservation due to the approximations made in  \cite{Nahrgang:2011mg}. It is numerically small. In figure  \ref{fig:equi_sourcetermtot} we also show the difference between $\Delta  E_{\rm diss}$ and $\Delta E_\sigma$ including the noise term. We identify this difference with $\Delta E_\xi$.

\section{Equilibration for a heat bath with reheating}\label{sec:equi_reheating}
In this section we investigate the influence of the energy conservation on the equilibration of the entire system. While the sigma field relaxes after a sudden temperature quench energy dissipates from the system to the heat bath. This in return changes the temperature of the quark fluid and the effective potential. Thus, the evolution of the sigma field itself is altered. In the last section we discussed the energy transfer between the sigma field and the heat bath. It has the two components $\Delta E_{\rm diss}$ and $\Delta E_\xi$. In the following we locally calculate $\Delta E_\sigma$ and add this to the local energy density of the heat bath given by (\ref{eq:lsm_eos2}). The new energy density is inverted to find the local temperature.

\begin{figure}
 \centering
 \subfigure[]{\label{fig:equi_avssigmareheatingfo}\includegraphics[width=0.45\textwidth]{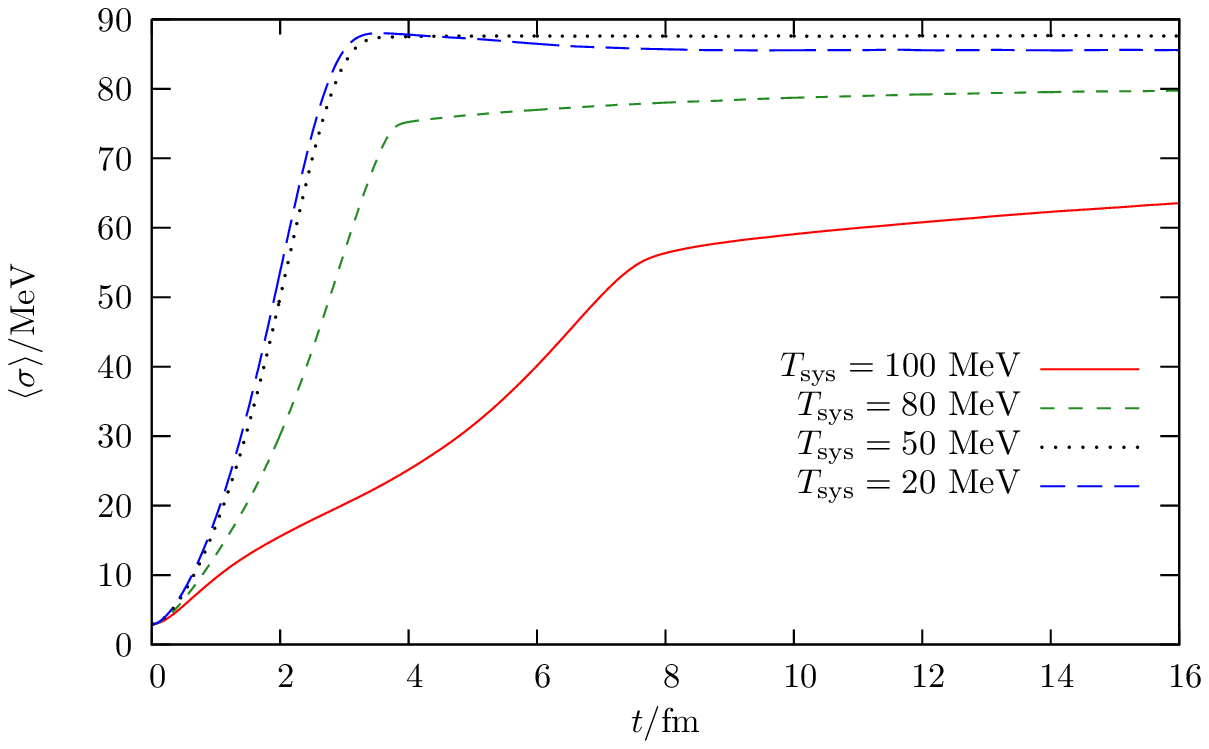}}
 \subfigure[]{\label{fig:equi_avssigmareheatingcp}\includegraphics[width=0.45\textwidth]{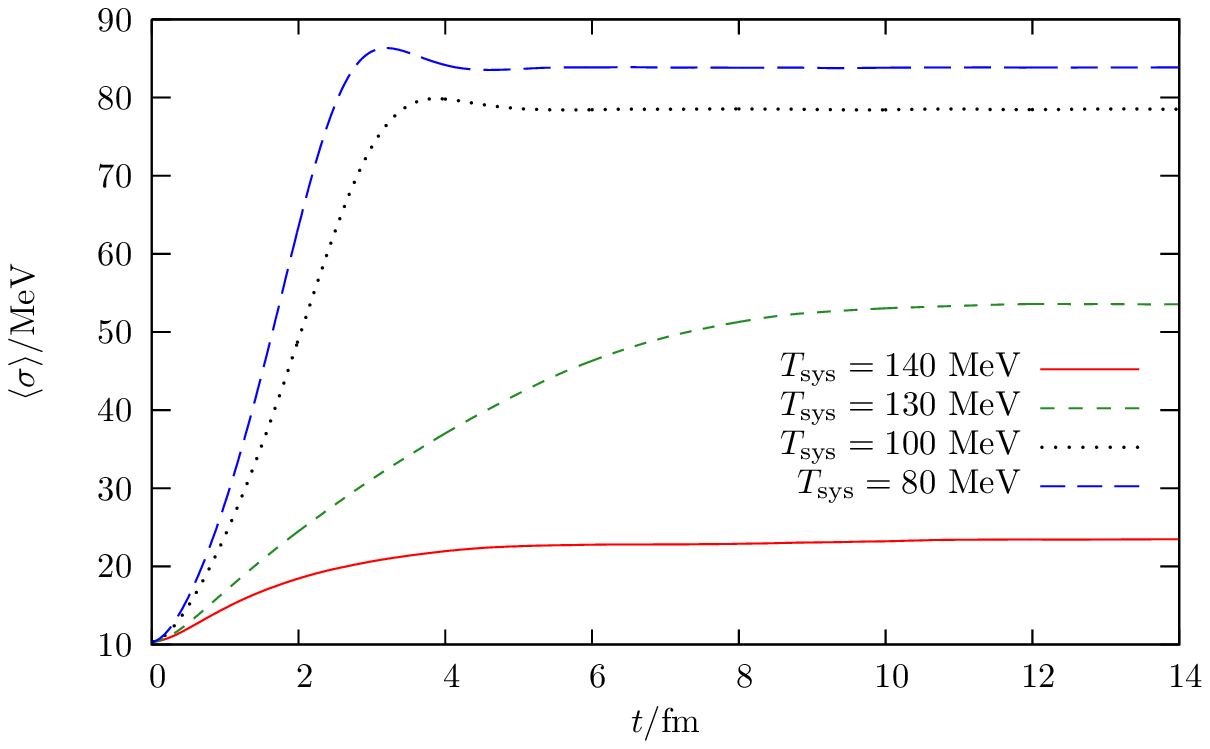}}
 \caption{Time evolution of the volume averaged sigma field in a scenario with a first order phase transition \subref{fig:equi_avssigmareheatingfo} and with a critial point \subref{fig:equi_avssigmareheatingcp} for different temperature quenches. The energy dissipation from the sigma field to the heat bath is taken into account.}

 \label{fig:equi_avssigmareheatingtot}
\end{figure}

\subsection{First order phase transition}
\begin{figure}
  \centering
 \subfigure[]{\label{fig:equi_avstempreheatingfo}\includegraphics[width=0.45\textwidth]{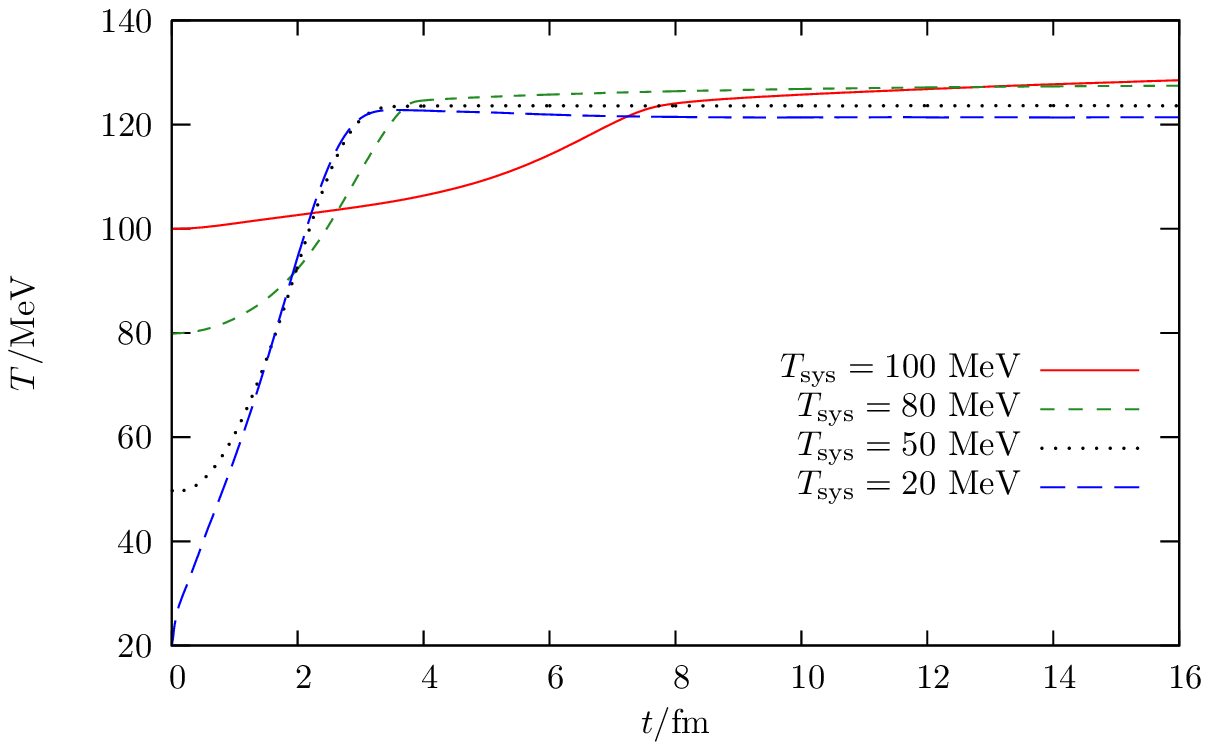}}
 \subfigure[]{\label{fig:equi_avstempreheatingcp}\includegraphics[width=0.45\textwidth]{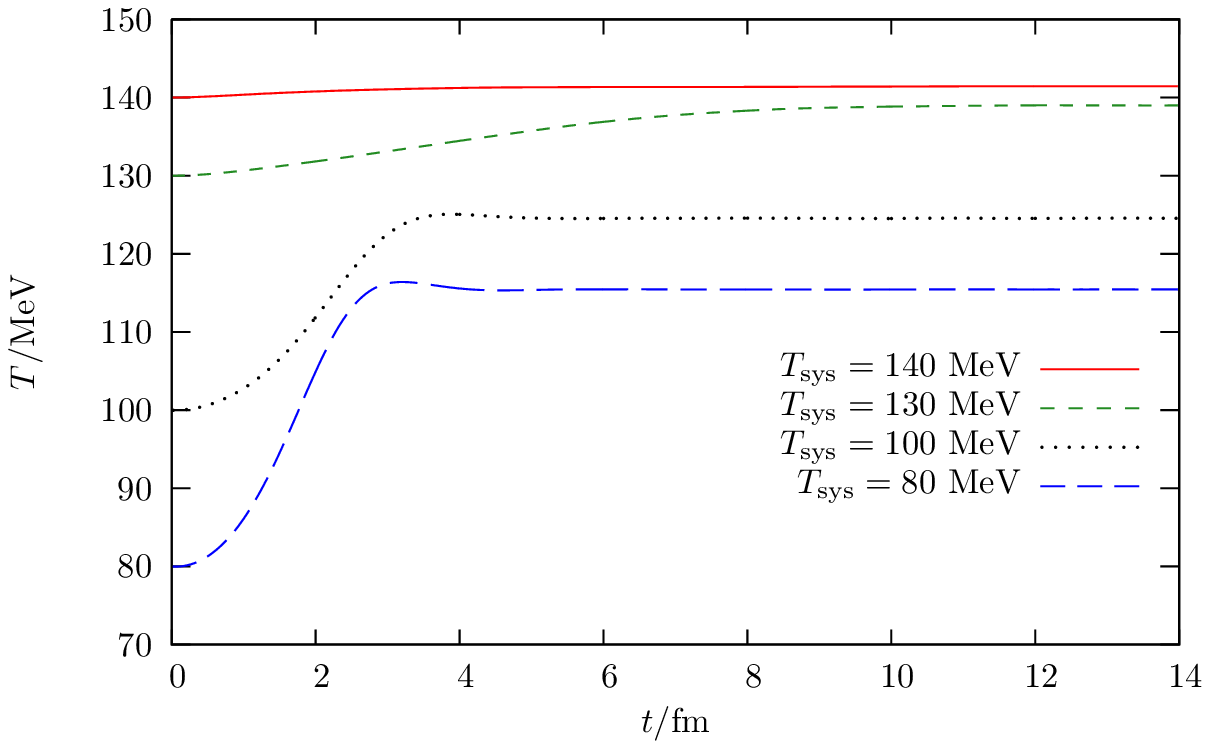}}
  \caption{Time evolution of the temperature in a scenario with a first order phase transition \subref{fig:equi_avstempreheatingfo} and with a critical point \subref{fig:equi_avstempreheatingcp} for different temperature quenches. The temperature is changed by the energy dissipation from the sigma field to the heat bath.}
  \label{fig:equi_avstempreheatingtot}
 \end{figure}

We present four results for scenarios with a first order phase transition. We quench from $T_{\rm ini}=160$~MeV to $T_{\rm sys}=100,\, 80,\, 50$ and $20$~MeV. During the relaxation of the volume averaged sigma field, see figure \ref{fig:equi_avssigmareheatingfo}, the average temperature increases rapidly to $T_{\rm fin}$, see figure \ref{fig:equi_avstempreheatingfo}. The exact values are shown in table \ref{table:equi_foreheating}. Three temperatures are above  $T_c$ and below or close to the upper spinodal temperature $T_{\rm sp}^{(2)}$, where the effective potential has two minima. The sigma field initially relaxes towards the vacuum value. This relaxation reheats the heat bath and causes an increase in the temperature to above $T_c$. Large parts of the sigma field now remain in the unstable low-temperature minimum. We see that including reheating the entire system does not equilibrate for these temperatures. Obviously, the reheating locally changes the effective potential such that it counteracts the relaxational process. Only for the very low temperature $T_{\rm sys}=20$~MeV, which is close to vacuum conditions, the final temperature is below $T_c$. Thus, the initial relaxation of the sigma field corresponds already to the equilibrium state at $T_{\rm fin}$.

\subsection{Critical point}
In a scenario with a critical point the effective potential has only one minimum for all temperatures. Therefore, we expect the entire system to equilibrate. We consider the following four temperature quenches from $T_{\rm ini}=160$~MeV to $T_{\rm sys}=140,\, 130,\, 100,\,{\rm and}\, 80$~MeV respectively. The corresponding volume averaged values are shown in table \ref{table:equi_cpreheating}. The volume averaged variances of both quantities are explicitly given. We clearly see that the entire system relaxes at a temperature $T_{\rm fin}$ and $\sigma_{\rm fin}\simeq\sigma_{\rm eq}(T=T_{\rm fin})$ in figures  \ref{fig:equi_avssigmareheatingcp} and \ref{fig:equi_avstempreheatingcp}. We observe that for a temperature quench to $T_{\rm sys}=130$~MeV the final temperature comes closest to the critical temperature $T_c=139.88$~MeV. As seen in figure \ref{fig:equi_avssigmareheatingcp} and figure \ref{fig:equi_avstempreheatingcp}, relaxation times are longest for this quench.

\begin{table}
\centering
 \subfigure[]{\begin{tabular}{| c | c | c | c |}
    \hline
    $T_{\rm sys}$/MeV & $T_{\rm fin}$/MeV & $\sigma_{\rm fin}$/MeV & $\sigma_{\rm eq}(T=T_{\rm fin})$/MeV\\ \hline
    100 & 131.89 & 69.06 & 5.00\\ 
    80 & 127.69 & 80.11 & 5.56\\
    50 & 123.59 & 87.57 & 6.25\\
    20 & 121.41 & 85.60 & 84.82\\
    \hline
  \end{tabular}\label{table:equi_foreheating} }
  \subfigure[]{ \begin{tabular}{| c | c | c | c |}
    \hline
    $T_{\rm sys}$/MeV & $T_{\rm fin}$/MeV & $\sigma_{\rm fin}$/MeV & $\sigma_{\rm eq}(T=T_{\rm fin})$/MeV\\ \hline
    $140$ & $141.42\pm 0.31$ & $23.33\pm 1.92$ & $24.37$\\ 
    $130$ & $138.96\pm 0.47$ & $53.43\pm 1.52$ & $54.52$\\
    $100$ & $124.53\pm 0.57$ & $78.46\pm 1.40$ & $78.60$\\
    $80$  & $115.44\pm 0.62$ & $83.82\pm 1.43$ & $83.90$\\
    \hline
  \end{tabular}\label{table:equi_cpreheating} }
\caption{Exact values for the relaxation of the volume averaged sigma field and the final temperatures for the different quenches for $T_{\rm ini}=160$~MeV to $T_{\rm sys}$. Here for a scenario with a first order phase transition in \subref{table:equi_foreheating} and with a critical point in \subref{table:equi_cpreheating}.}
\label{table:equi_totreheating}
\end{table}

\section{Summary and Outlook}
We have studied the dynamics of the sigma field given by the Langevin equation (\ref{eq:equi_langevineq}) with the damping (\ref{eq:equi_damping1}) and (\ref{eq:equi_damping2}) and the noise correlator (\ref{eq:sc_noisecorrelation}). It leads to the relaxation of the sigma field with a static isothermal heat bath. Including reheating of the heat bath we find full relaxational dynamics only for a scenario with a critical point. For a first order phase transition the system stays in the low-temperature minimum, which due to reheating becomes the unstable minimum at the final temperature. At the critical point we observed longest relaxation times which is in accordance with critical slowing down.
In future work we will include the fluid dynamic expansion of the heat bath and thus study the full nonequilibrium chiral fluid dynamics.

 The authors thank Carsten Greiner for fruitful discussions.
 This work was supported by the Hessian LOEWE initiative Helmholtz International Center for FAIR. M.~Nahrgang acknowledges financial support from the Stiftung Polytechnische Gesellschaft Frankfurt.

\end{document}